\documentclass[british,english,aps, paper, twocolumn]{revtex4}
\usepackage[T1]{fontenc}
\usepackage[utf8]{inputenc}
\usepackage{listings}
\usepackage{babel}
\usepackage{array}
\usepackage{multirow}
\usepackage{graphicx}
\usepackage[unicode=true,pdfusetitle,
 bookmarks=true,bookmarksnumbered=false,bookmarksopen=false,
 breaklinks=false,pdfborder={0 0 1},backref=false,colorlinks=false]
 {hyperref}

\makeatletter

\providecommand{\tabularnewline}{\\}

\newcommand{\lyxdeleted}[3]{}

\@ifundefined{textcolor}{}
{%
 \definecolor{BLACK}{gray}{0}
 \definecolor{WHITE}{gray}{1}
 \definecolor{RED}{rgb}{1,0,0}
 \definecolor{GREEN}{rgb}{0,1,0}
 \definecolor{BLUE}{rgb}{0,0,1}
 \definecolor{CYAN}{cmyk}{1,0,0,0}
 \definecolor{MAGENTA}{cmyk}{0,1,0,0}
 \definecolor{YELLOW}{cmyk}{0,0,1,0}
}

\usepackage{lmodern}  

\usepackage{xcolor}
\usepackage{soul}
\providecolor{lyxadded}{rgb}{0,0,1}
\providecolor{lyxdeleted}{rgb}{1,0,0}

\DeclareRobustCommand{\lyxdeleted}[3]{{\color{lyxdeleted}\st{#3}}}

\makeatother

\begin{document}

\title{Foucault's method in new settings}

\author{Z. Vörös, G. Weihs}

\affiliation{Department of Experimental Physics, University of Innsbruck, Technikerstraße
25/d, Innsbruck, A-6020 Austria}
\begin{abstract}
In this paper, we introduce two simple and inexpensive versions of
the well-known Foucault method for measuring the speed of light. In
a footprint of just 20\,cm by 270\,cm with readily available laboratory
items and a \foreignlanguage{british}{webcam}, we obtained $c=296720\pm3000$\,km/s,
and $c=302295\pm3000$\,km/s, respectively, both within less than
a per cent of the defined value. The experiment also prepares students
to work with large amounts of data. 
\end{abstract}
\maketitle

\section{Introduction}

In recent years, we could witness a substantial paradigm shift in
sciences: relationships emerge from vast amounts of collected data
(sometimes dubbed \textit{big data}), and not just a few measurements.
Notable examples are customer recommendation systems of Amazon, eBay
and similar retailers, or the data streams of the Large Hadron Collider
or the Square \foreignlanguage{british}{Kilometre} Array. This trend
is expected to continue in the foreseeable future, especially, with
the advent of \foreignlanguage{british}{always-online} mobile devices
capable of continuously collecting and transmitting all kinds of data. 

However, it also seems that science education does not keep up with
the pace of progress in data collection and analysis capabilities,
and that many introductory or even advanced level laboratory experiments
are still conducted with hand-held stopwatches, weights, mercury thermometers,
and microscopes with engraved scales. This approach has at least three
inherent problems. The first is that it teaches students how experiments
were conducted two centuries ago, but does not tell them how to do
them now. Second, the amount of data that can be collected in this
way is limited, and inaccurate. This also means that statistical evaluation
of the results is constrained to a handful of data points. Finally,
by the very nature of the required specialized setups, these experiments
are expensive, and students have to operate within the spatial and
temporal confines of the laboratory course. We are convinced that
one cannot underestimate the pedagogical benefits of pursuing science
on the kitchen sink: when one can accurately measure something relevant
(such as, a fundamental constant) with easily available and cheap
everyday items, and without reference to a dedicated laboratory. It
is a very fortunate coincidence that in our times, everyday items
are digital gadgets capable of measuring all kinds of physical quantities,
e.g., distance, temperature, acceleration, magnetic fields, light
intensity, frequency, time etc., and that the demand for high quality
in user experience makes it possible to deliver unprecedented accuracy. 

At the same time, we also recognize the pedagogical value of discussing
how experiments were conducted in the past and that it would be an
irreparable loss not to show what could be achieved with devices that
we would now consider rudimentary. What we would like to demonstrate
in this paper is that it is not necessary to regard the above-mentioned
two subjects, historical perspective, and progress in measurement
capabilities as disjoint. There are ways of showing the beauty and
ingenuity of past experiments, while reaping the many benefits of
modern technologies in terms of measurement time, accuracy, cost,
or the volume of data. 

The example that we take is the measurement of the speed of light,
which is one of the fundamental physical constants. Strictly speaking,
our example is pathological in the sense that the international meter
is defined by the help of the speed of light and the international
standard of time, and not the other way around. However, first, till
1983 (i.e., in Foucault's life), length and time were defined and
the speed of light was the derived quantity, second, this fact does
not reduce the didactic value of the experiment itself. We would like
to emphasize that while the evaluation of the measurements requires
some data processing, this fact does definitely not qualify it as
a \emph{big data} exercise.

The paper is organized as follows. In the next two sections, we outline
the historical and theoretical background and derive the expression
for $c$. In Section IV., we introduce our experimental setup and
the critical components, Section V. contains a detailed discussion
of our results, while Section VI. is devoted to a thorough analysis
of various systematic errors. In the appendix, we present a couple
of MATLAB (The Mathworks, Inc.) snippets that can be used to evaluate
measurement data.

\section{Historical background}

That the speed of light, $c$, is finite was already conjectured by
Galileo in the XVII. century, though, his experimental apparatus at
the time prevented him from giving even an order-of-magnitude estimate
for the value. Since then, various methods have been developed. 

It was first Huygens, who, based on the astronomical measurements
of Rømer in 1676 on the entry into and exit from eclipses of Jupiter's
moons, could provide a lower bound of about 200 000\,km/s. The same
measurements, repeated with higher accuracy by Delambre in 1809, yielded
304 000\,km/s, astonishingly close to the true value. Another astronomical
method, the aberration of light, was discovered by Bradley in 1729,
with the result of about 296 000\,km/s. 

Later, it was realized that in Maxwell's theory, the speed of light
is linked to fundamental electromagnetic constants through the relation
$\epsilon_{0}\mu_{0}=c^{-2}$, and therefore, by measuring the vacuum
permittivity $\epsilon_{0}$, and the vacuum permeability $\mu_{0}$,
it is possible to indirectly infer the value of $c$ \cite{Clark1956,Clark2001}.

It is also to be noted that, if the frequency $f$ of electromagnetic
radiation is known, and the wavelength $\lambda$ can be measured,
then by dint of the relation $c=f\lambda$, $c$ can be indirectly
determined. This is the basis of measurements of interferometric methods
\cite{Belich1996,Lahaye2012}, and of cavity resonance methods \cite{DOrazio2010}. 

Finally, there are several methods that measure the time of flight
in terrestrial settings. One of them is the Foucault method that we
discuss in more detail in the next section \cite{Dillman1964,Feagin1979,Morrison1980,Brody2010},
while with the advent of high-speed electronics, it is now possible
to directly measure the delay in the arrival of short optical pulses
as the distance between the emitter and receiver is increased \cite{Rogers1969,Deblaquiere1991,Aoki2008,Ronzani2008}.
This latter method is the simplest of all, but it definitely lacks
the elegance of the others. 

The interested reader can find a more detailed survey of various measurement
methods and their significance in \cite{Bates1988}.

\section{Theoretical background}

Foucault's is one of the simplest methods of measuring the speed of
light on Earth, and it falls into the category of time of flight measurements.
It is based on the observation that, if a light beam bounces off a
moving mirror twice, the mirror will have moved by a small amount
by the time it is hit by the beam the second time, and this movement
results in a small displacement of the reflected beam. It is this
displacement that is to be measured, and from where the speed of light
is to be inferred. In this particular instance, the mirror is rotating,
and the rotation angle between the two events can simply be related
to the time that was required for the round trip. The speed of light
can be obtained from the measured displacement, and the length of
the round-trip path. Due to its conceptual simplicity, this is perhaps
the most popular method in student laboratories. 

\begin{figure}[h]
\includegraphics[width=0.98\columnwidth]{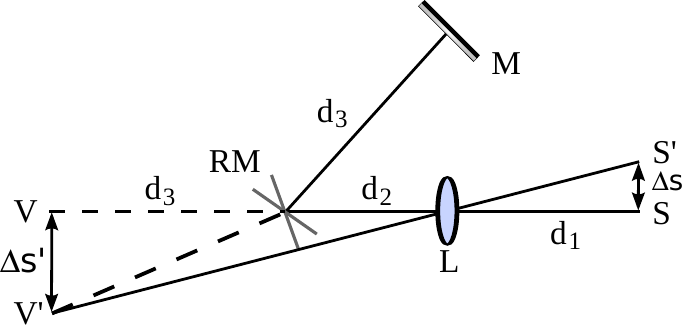}

\caption{The concept of the experiment. $RM$ is the rotating mirror, $M$
is the end reflector, and $L$ is a lens. $S,S'$ are the light source,
and its image, respectively, while $V$, and $V'$ are the virtual
images of $S$, and $S'$.}

\label{fig:concept}
\end{figure}

To be more specific, let us take the simplified experimental setup
shown in Fig.\,\ref{fig:concept}. A point source emitting light
is located at point $S$, at a distance $d_{1}$ from the lens $L$.
The source's light is reflected by the rotating mirror $RM$ (at a
distance of $d_{2}$ from the lens, and at this point, stationary)
and its image is created at the position of the end mirror $M$, which
is at a distance of $d{}_{3}$ from the rotating mirror, and is normal
to the in-coming light. This also means that the light reflected by
$M$ is focused on $S$ again. In the absence of the rotating mirror,
the image of $S$ would be at $V$. 

Now, let us assume that in the time $\Delta t$ the light traverses
the distance between $RM$ and $M$ in both directions, the rotating
mirror turns by an amount $\omega\Delta t$, where $\omega$ is the
angular velocity, and $\Delta t=2d_{3}/c$. This rotation displaces
the virtual image $V$ to $V'$, where the distance between these
two points is simply $\Delta s'=2\omega\Delta t\cdot d{}_{3}$. The
factor of $2$ is a result of the reflection on $RM$: upon reflection,
all angles change by a factor of $2$. The image of the virtual point
$V'$ is mapped by the lens to the point $S'$, and using the two
similar triangles formed by $V,V'$, and the lens, and $S,S'$, and
the lens, respectively, we conclude that the distance between $S$
and $S'$ is 
\[
\Delta s=\frac{d_{1}}{d_{2}+d_{3}}\Delta s'=2\omega\Delta t\frac{d_{1}d_{3}}{d_{2}+d_{3}}=\frac{4d_{1}d_{3}^{2}}{d_{2}+d_{3}}\frac{\omega}{c}\ ,
\]
i.e., the speed of light is 
\begin{equation}
c=\frac{4d_{1}d_{3}^{2}}{d_{2}+d_{3}}\frac{\omega}{\Delta s}\ .\label{eq:c_final}
\end{equation}
Given $d_{1,2,3}$, the speed of light can be gotten by measuring
the displacement $\Delta s$ for a given angular speed. In principle,
to determine $c$, a single measurement point is enough, but as we
will see later, by measuring $\Delta s$ as a function of $\omega$,
and taking the slope of the linear dependence, it is not necessary
to find the reference position at $\omega=0$. Re-arranging Eq.(\ref{eq:c_final})
yields 
\begin{equation}
c_{0}=\frac{4d_{1}d_{3}^{2}}{d_{2}+d_{3}}\left(\frac{d\Delta s}{d\omega}\right)^{-1}\ .\label{eq:c_diffs}
\end{equation}
In Section VI., we will show that the errors are negligible, if the
lens is not positioned perfectly, and the image of $S$ is not formed
at $M$. In the formula above, $c_{0}$ indicates that these errors
are not yet taken into account.

\section{Experimental setup}

\begin{figure}[h]
\includegraphics[width=0.98\columnwidth]{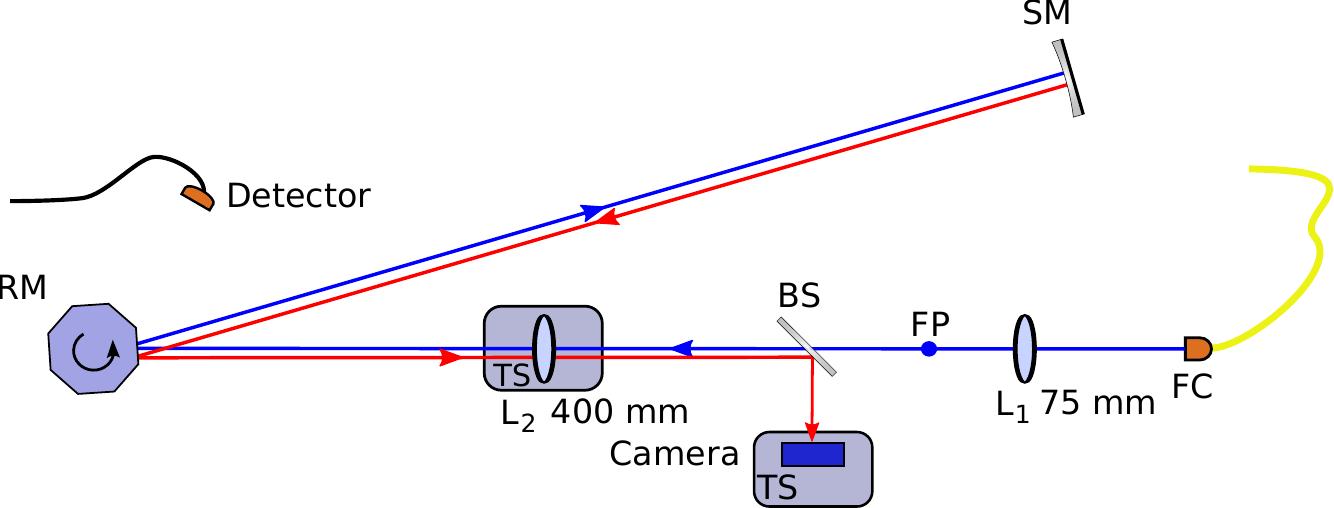}

\caption{Experimental setup Setup 1. $RM$, $FM$, $SM$ are the rotating,
folding, and back reflector mirrors, respectively, $L_{1}$, $L_{2}$
are lenses of focal length 75 and 400\,mm, respectively, $FP$ is
the focal point of $L_{1}$, $FC$ is the fibre collimator, $BS$
is the beamsplitter, and $TS$ are translation stages. Dimensions
are given in the text.}

\label{fig:experimental_setup}
\end{figure}

Fig.\,\ref{fig:experimental_setup} displays our first experimental
setup. Laser light from a standard fibre fault locator (OZ Optics,
FODL-43S-635-1) emitting at a wavelength of 635\,nm is transmitted
through a single-mode fibre, and collimated by a short focal length
fibre collimator (Thorlabs F240FC-B). While it is not absolutely necessary,
by passing the light through the single-mode fibre patch cable (Thorlabs
P1-630A-FC-1), we begin with a perfect Gaussian beam. It is worth
noting that the fault locator can be replaced by an inexpensive laser
pointer. 

The collimated beam is then lead through a telescope consisting of
two lenses of focal lengths 75\,mm ($L_{1}$), and 400\,mm ($L_{2}$),
respectively. The telescope is misaligned slightly in the longitudinal
direction (the distance between the two lenses is larger than 475\,mm),
so that the beam leaving is not collimated any more, but, after being
reflected on the rotating mirror $RM$, is focused on a spherical
mirror $SM$, which acts as the back reflector. The rotating mirror
is located at a distance of 1630\,mm from the 400-mm lens, while
the back reflector with a radius of curvature of 4000\,mm is positioned
at a distance of 4830\,mm from the rotating mirror. Distances were
measured with a tape measure. In order to reduce the overall size
of the setup, the 4830-mm path was folded by the insertion of a flat
mirror (not shown) between $RM$, and $SM$, $FM$. We should also
note that since the spherical mirror is not involved in the imaging,
it can be replaced by a flat mirror.

For monitoring the rotation, we also placed a standard silicon photodiode
(Thorlabls PD136A) close to the rotating mirror: when rotating, the
mirror diverts the laser light to the diode 8 times per revolution,
thereby, producing a well-defined potential spike that can conveniently
be recorded on an oscilloscope.

The light reflected by the spherical mirror travels along the same
path, except that it is diverted to a \foreignlanguage{british}{webcam}
(Logitech C310) by a pellicle beam splitter ($BS$, Thorlabs BP150)
positioned to the left of $FP$, which is the focal point of $L_{1}$.
The small lens of the webcam has to be removed before use, so that
no extra imaging element is introduced. In the original version of
the experiment, instead of a camera, a microscope is used to measure
the displacement of the beam. However, given the finite size of the
focal spot, this also entails that large distances have to be employed
in order to realize measurable displacements. The application of the
camera not only makes data collection more convenient, but it also
implies that the physical size of the setup can considerably be reduced.
We would like to point out that the use of a webcam in the context
of speed of light measurements was discussed in an interferometric
setting in \cite{Lahaye2012}.

\begin{figure}[h]
\includegraphics[width=0.98\columnwidth]{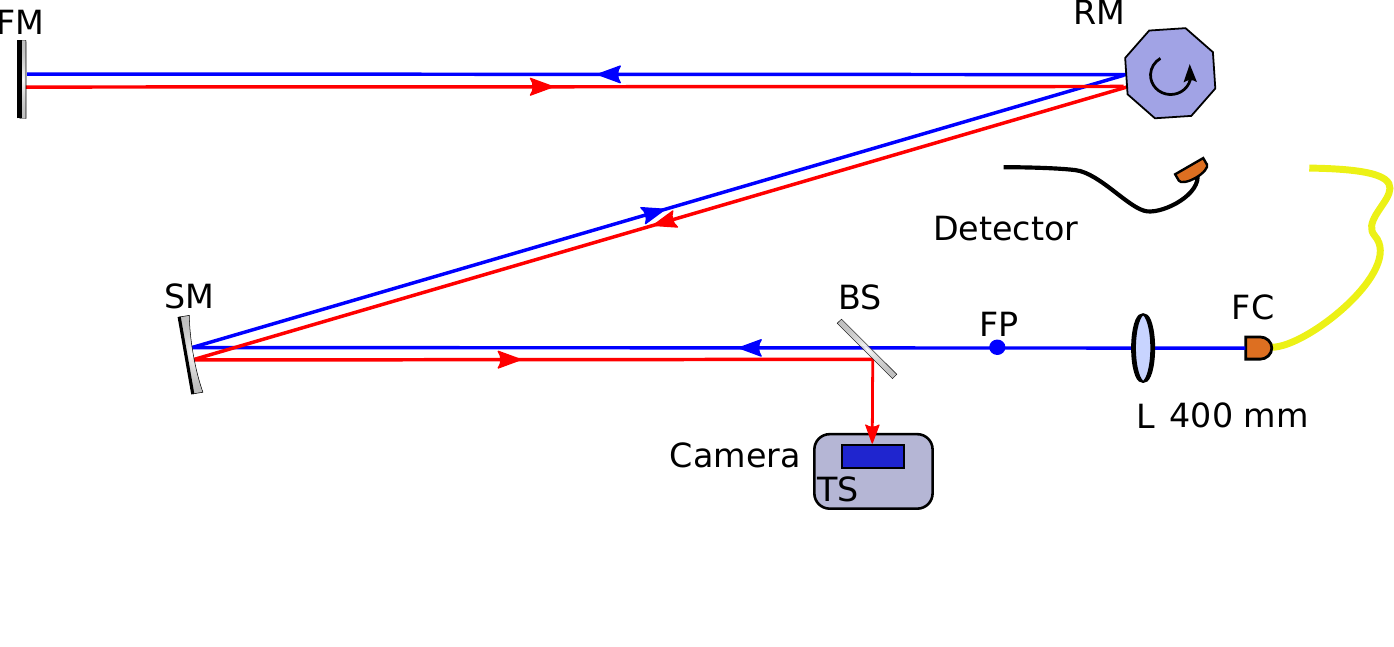}

\caption{Experimental setup, Setup 2. $RM$, $FM$, and $SM$ are the rotating,
flat, and spherical mirrors, respectively, $L$ is a lens of focal
length 400\,mm, $FC$ is the fibre collimator, $BS$ is the beamsplitter,
and $TS$ are translation stages. Dimensions as indicated in the text.}

\label{fig:experimental_setup2}
\end{figure}

Our second setup, Setup 2, is shown in Fig.\,\ref{fig:experimental_setup2}.
The light of the fault locator is focused by a lens of focal length
$400\,\mathrm{mm}$ onto $FP$, from where it reaches the spherical
mirror $SM$ with a focal length of 2\,m. The spherical mirror is
$4060\,\mathrm{mm}$ away from $FP$, and is tilted slightly, so that
the light is reflected off the rotating mirror $RM$, located $730\,\mathrm{mm}$
away from $SM$, and finally $FM$, located $3260\,\mathrm{mm}$ away
from $RM$. The lengths in the setup are chosen in such a way that
the light is focused on the flat end reflector, $FM$, although, as
will be discussed in Section VI., small longitudinal misalignments
do not influence the results in any significant way. As in the first
setup, folding mirrors were used between $SM$, and $RM$, and between
$RM$, and $FM$.

The two setups are conceptually the same: the only difference between
them is that the imaging element in the first case is a lens, while
in the other case, it is a spherical mirror.

\begin{figure}[h]
\includegraphics[width=0.98\columnwidth]{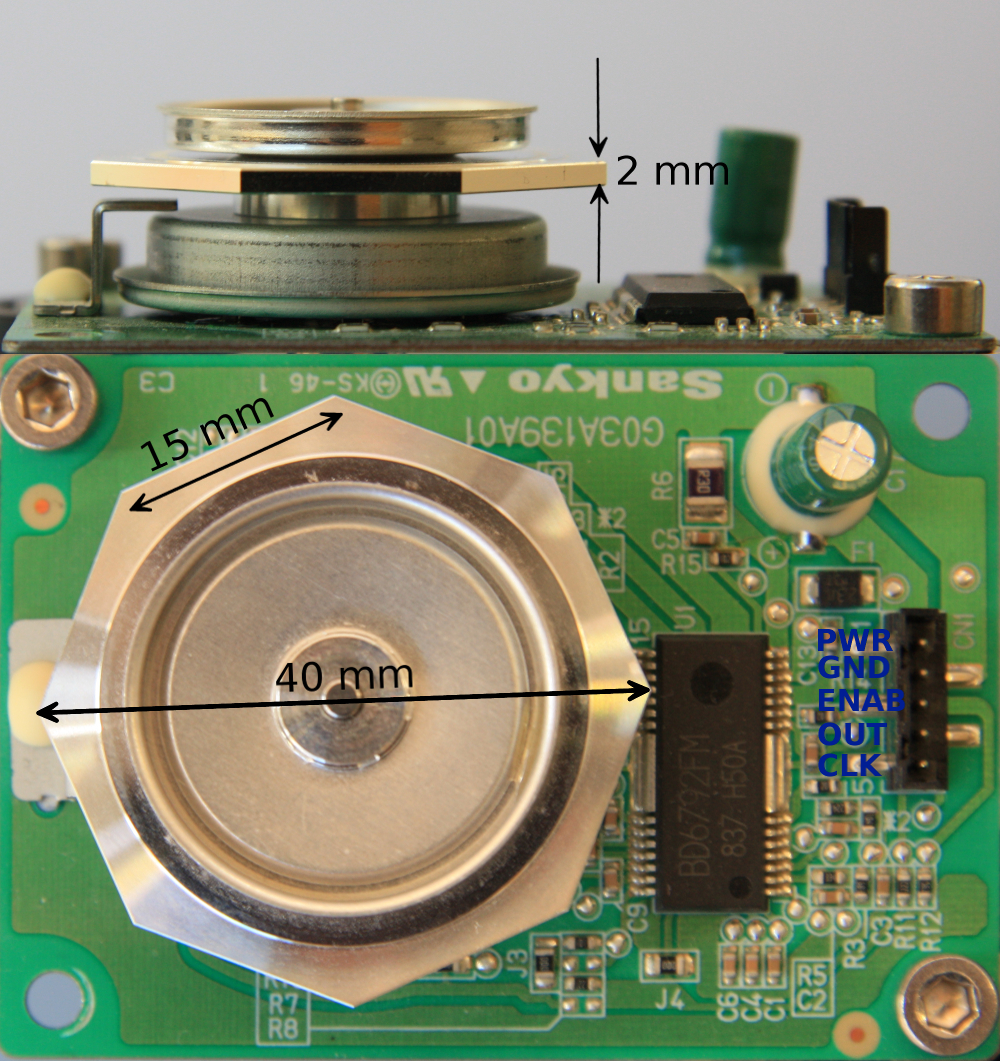}

\caption{Rotating printer mirror, side view (top), and top view (bottom). The
30-pin integrated circuit contains the motor driver (BD6792FM from
Rohm Semiconductors) with the built-in PLL. Control pins are labeled
in blue.}

\label{fig:8mirror}
\end{figure}

As the rotating reflector, we employed an octagonal printer mirror
scavenged from a faulty printer, shown in Fig.\,\ref{fig:8mirror}.
(This part can also be purchased separately. A possible alternative
is a barcode reader with a revolving mirror.) Laser printers utilize
a focused laser beam to locally discharge a positively pre-charged
cylindrical drum, that is itself rotating around its own axis. The
octagonal (sometimes quadratic, or hexagonal) mirror is used for scanning
the laser beam along the axis of the drum, thereby creating an accurate
time-to-two-dimensional mapping on the drum's surface. In order to
achieve high spatial accuracy, both the drum and the rotating mirror
have to revolve at a constant speed. Stabilization of the rotation
frequency is achieved by means of phase-locked loops (PLL), in which
an external clock signal is locked to the signal of a magnetic field
transducer measuring the temporal variations of the field of a constant
magnet moving with the axle of the motor. This also means that, within
limits, the rotation speed can be set by adjusting the clock signal
that is fed into the PLL loop. Fig.\,\ref{fig:8mirror} also indicates
the connections of the mirror assembly: $PWR$ (pin 1) is the power
line, whose potential can be anything between +18, and +36\,V, $GND$
(pin 2) is ground, $ENAB$ (pin 3) is the active-low motor enable
pin (this should be tied to ground), while $CLK$ (pin 5) is the clock
line, which takes TTL pulses with frequencies between around 300,
and 6000\,Hz. Pin 4 is an output connected to the magnetic field
transducer, and can be used for monitoring the rotation. 

The advantages of the mirror assembly are that first, the mirror is
monolithic, therefore, it is safe to operate: no pieces can break
off at high speeds. Second, the control electronics makes it possible
to adjust the speed by setting the frequency of the clock signal from
a simple function generator, and that there is a well-defined linear
relationship between the rotation speed and the clock frequency.

Initial alignment of the setup is performed when the mirror is stopped
(the enable line is high). First, all mirrors are placed to their
respective positions, and $FC$ is aligned such that the collimated
laser beam can travel to the end mirror, $SM$. Then $L_{1}$ is inserted
in such a way that the diverging laser light still reaches both $RM$,
and $SM$. After this, $L_{2}$ is inserted in the path, and is moved
along the optical axis till the size of the light spot reaches its
minimum on $SM$. When this is achieved, the beamsplitter, $BS$,
has to be placed on the left hand side of the focal point of $L_{1}$,
at a distance of about 5-7\,cm from the focal spot, $FP$. With the
tip-tilt control knobs of the mirror holder, $SM$ has now to be aligned
so that the light is reflected back to the laser. At this point, the
reflected beam should be focused on $FP$. Finally, the camera has
to be placed in the diverted focus of the back-reflected beam. Great
care has to be taken to make sure that the camera's plane is as perpendicular
to the laser beam as possible: failure to do so will results in a
systematic error, which leads to higher speeds of light. For a thorough
discussion on this, see Section VI.

\section{Experimental results}

As can be inferred from Eq.(\ref{eq:c_diffs}), in order to determine
the speed of light, one has to measure $d_{1,2,3}$, the angular frequency
$\omega$, and the displacement $\Delta s$. The measurement can be
done in the same way in both setups, and the steps are as follows.
First, one has to determine the rotation speed as a function of the
clock frequency. Next, the pixel size of the camera has to be measured.
This step amounts to calibrating a ruler. Then the displacement of
the image on the camera has to be measured at various clock frequencies
(this step involves fitting to the camera images), and by using the
pixel size, this displacement has to be converted to physical units.
Finally, the slope of the displacement-frequency relationship has
to be determined, and inserted in Eq.(\ref{eq:c_diffs}).

\begin{figure}[h]
\includegraphics[width=1\columnwidth]{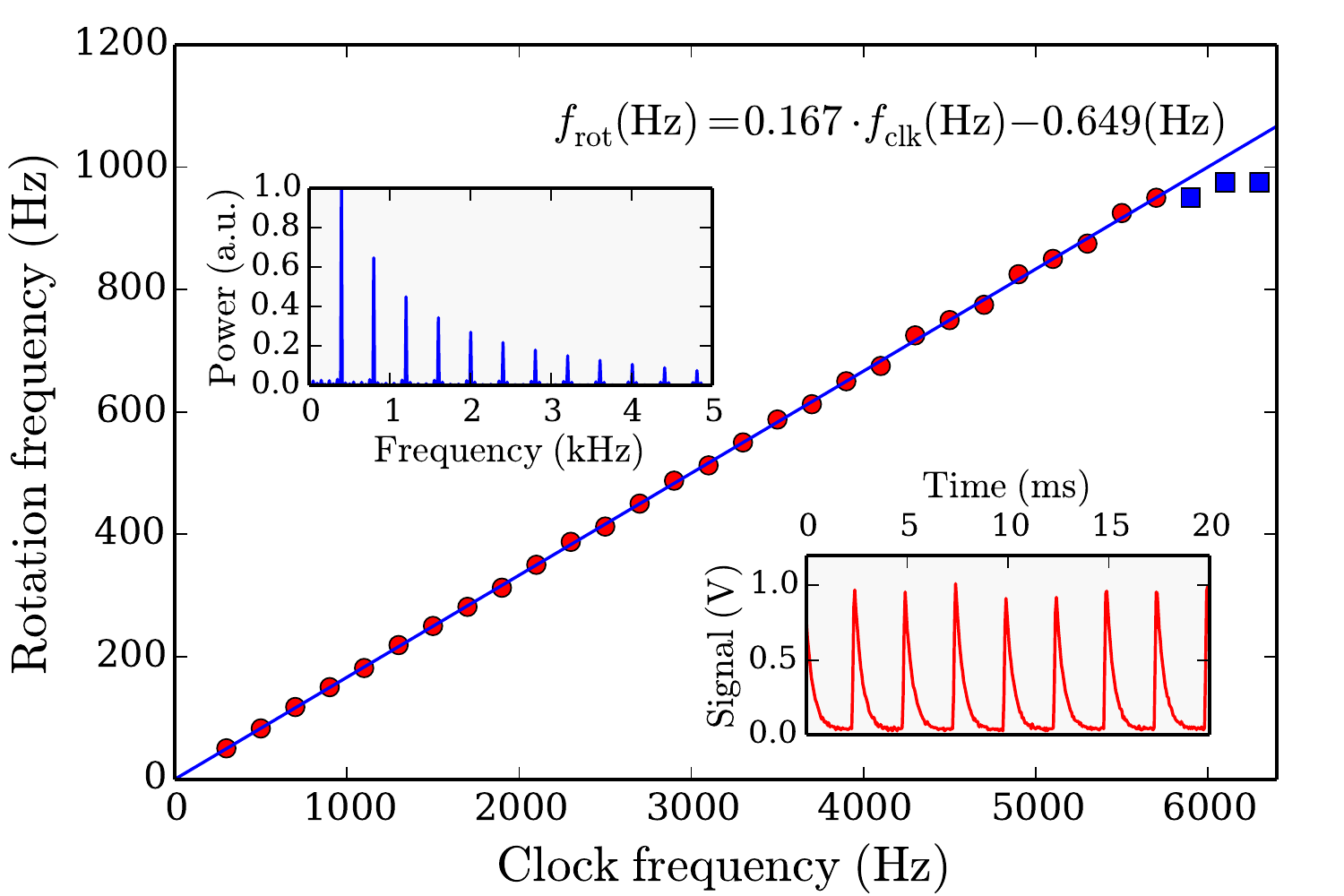}

\caption{Rotation speed as a function of clock frequency. The inset in the
lower right corner shows a typical time trace on the detector with
its Fourier spectrum in the upper left. The clock frequency for this
trace is 300 Hz. The parameters of a linear fit are displayed in the
figure. The last three data points (blue squares) were excluded. The
error in the slope is 0.00054. }

\label{fig:rpm_calibration}
\end{figure}

The rotation speed can be deduced from the time traces of the photodiode,
either by simply measuring the time difference between an integer
number of maxima, or recording the potential values, taking the Fourier
transform, and identifying the strongest frequency component. Given
a high enough number of samples, the two methods deliver the same
results. In Fig.\,\ref{fig:rpm_calibration}, we show the measured
rotation speed as a function of the clock frequency, with a typical
time trace of the detector signal on an oscilloscope, and its Fourier
transform. The period can clearly be resolved from either the signal,
or its Fourier spectrum. Note that at high clock rates, the rotation
speed saturates. For this reason, we excluded the last 3 points from
the linear fit, from which we deduced the relationship $f_{\mathrm{rot}}(Hz)=(0.167\pm0.00054)\cdot f_{\mathrm{clk}}(\mathrm{Hz)}-0.649\,\mathrm{(Hz)}$.
The error of the fit is approximately 0.3\%. Given the precision (in
the ppm range) of frequency standards used in modern pulse generators,
and the stability of phase-locked loops used in laser printers, we
ascribe the error to our way of determining the frequency from the
Fourier transform of the time trace. Also note that, since the rotating
mirror has 8 facets, the actual rotation speed is only 1/8 of what
the detector signal indicates.

It is worth pointing out that, given the order of magnitude of the
rotation speed, in the absence of an oscilloscope, these frequencies
can easily be measured by means of a smart phone. All one has to do
is to convert the electrical signal of the photodiode to sound by
amplifying it, and connecting it to a speaker, and then record the
sound through the microphone. There are countless applications that
can take and display the Fourier transform of the microphone input.
Likewise, the clock signal can be generated by a suitable waveform
applied to the phone's speaker.

\begin{figure}[h]
\includegraphics[width=1\columnwidth]{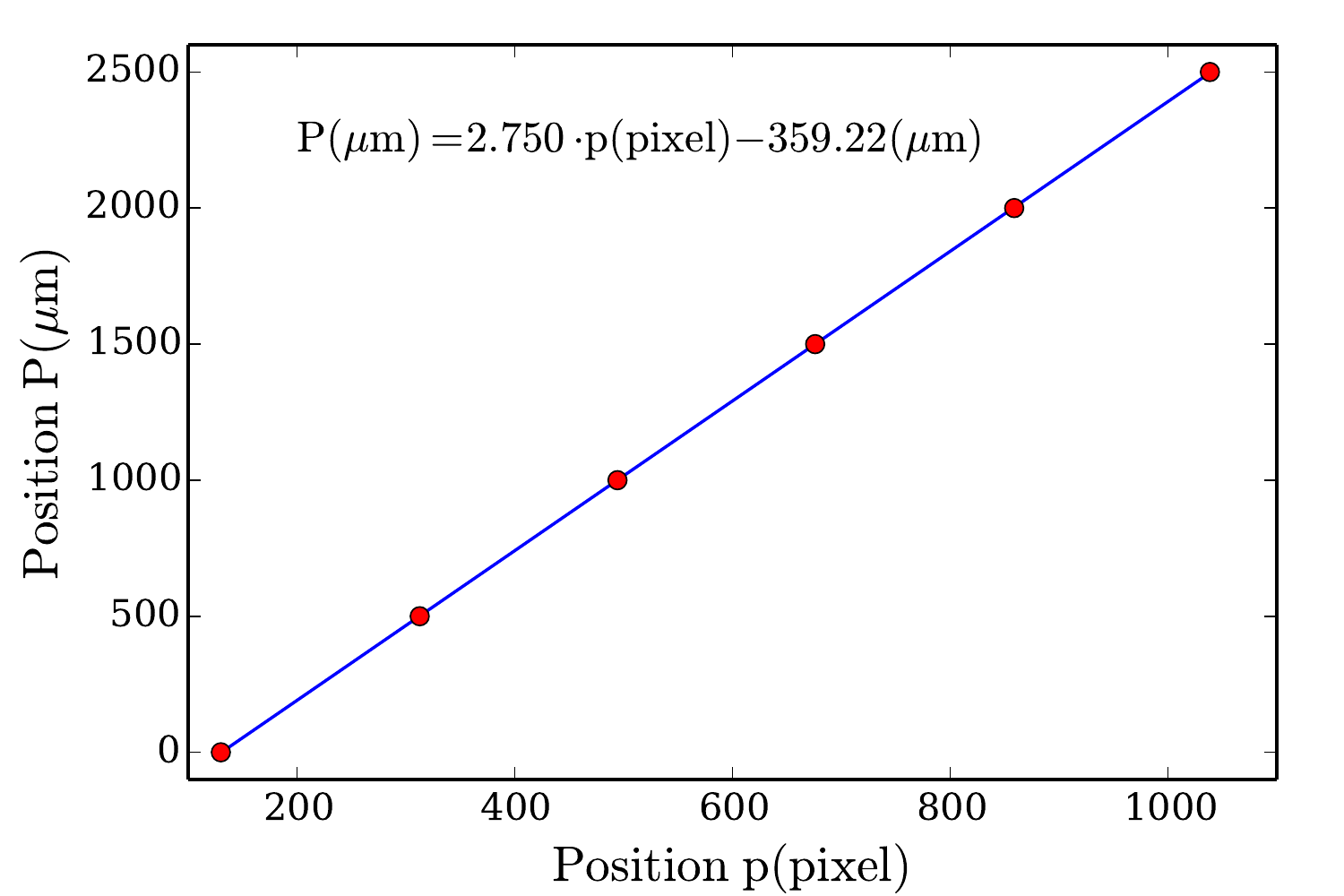}

\caption{Measurement of the pixel size. The statistical errors in both the
dependent and independent variables are too small to be visible. The
parameters of the linear fit are indicated in the figure. }

\label{fig:ccd_calibration}
\end{figure}

\begin{figure}[h]
\includegraphics[width=1\columnwidth]{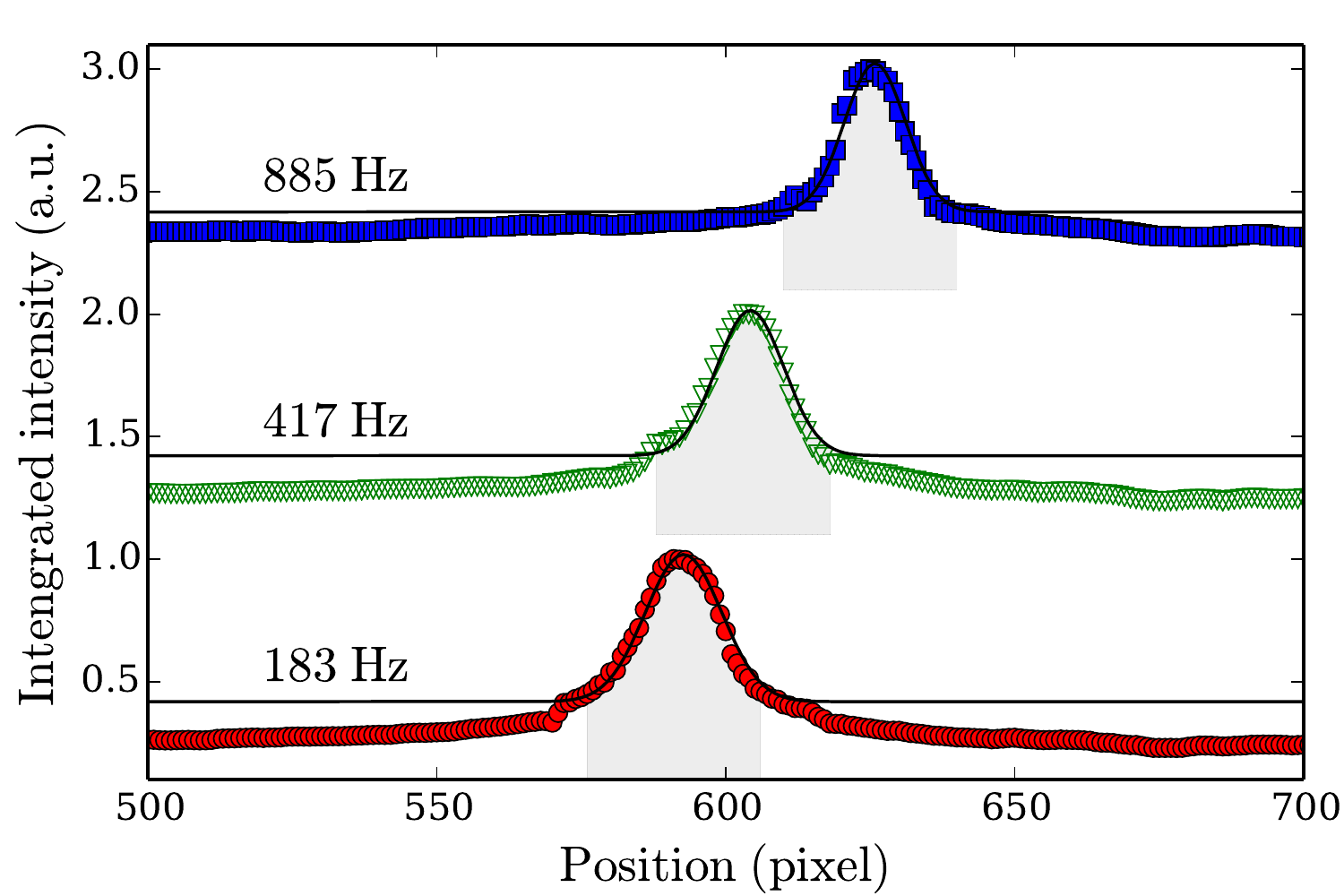}

\caption{Typical profiles taken in Setup 1 at a rotation frequency of 183 (solid
red circle), 417 (empty green triangle), and 885\,Hz (solid blue
square), respectively. The domain of Gaussian fits is indicated by
the shaded gray regions, while the solid black lines are the fits.}

\label{fig:profiles}
\end{figure}

In order to convert the pixel positions into physical distance, we
have to calibrate the CCD camera. In other words, we have to measure
the pixel size. For this procedure, we stopped the rotating mirror,
and shifted the camera by an amount indicated by the micrometer screw
on the translation stage. The data points are plotted in Fig.\,\ref{fig:ccd_calibration}
in conjunction with a linear fit, which gives a pixel size of $2.75\,\mu\mathrm{m}$.
This is also the value given by the manufacturer. By the help of this
measurement, one can also ascertain that the translation axis is parallel
to the camera's plane, because if that is not the case, then the width
of the profiles changes as the camera is shifted. As shown below (see
e.g., Fig.\,\ref{fig:profiles}), the centre of the nearly Gaussian
profiles can be obtained with sub-pixel accuracy. If we take half
of the smallest micrometer division ($10\,\mu\mathrm{m}$) as the
error in position, this procedure incurs an overall error of less
than one fifth of a per cent. 

\begin{figure}[h]
\includegraphics[width=1\columnwidth]{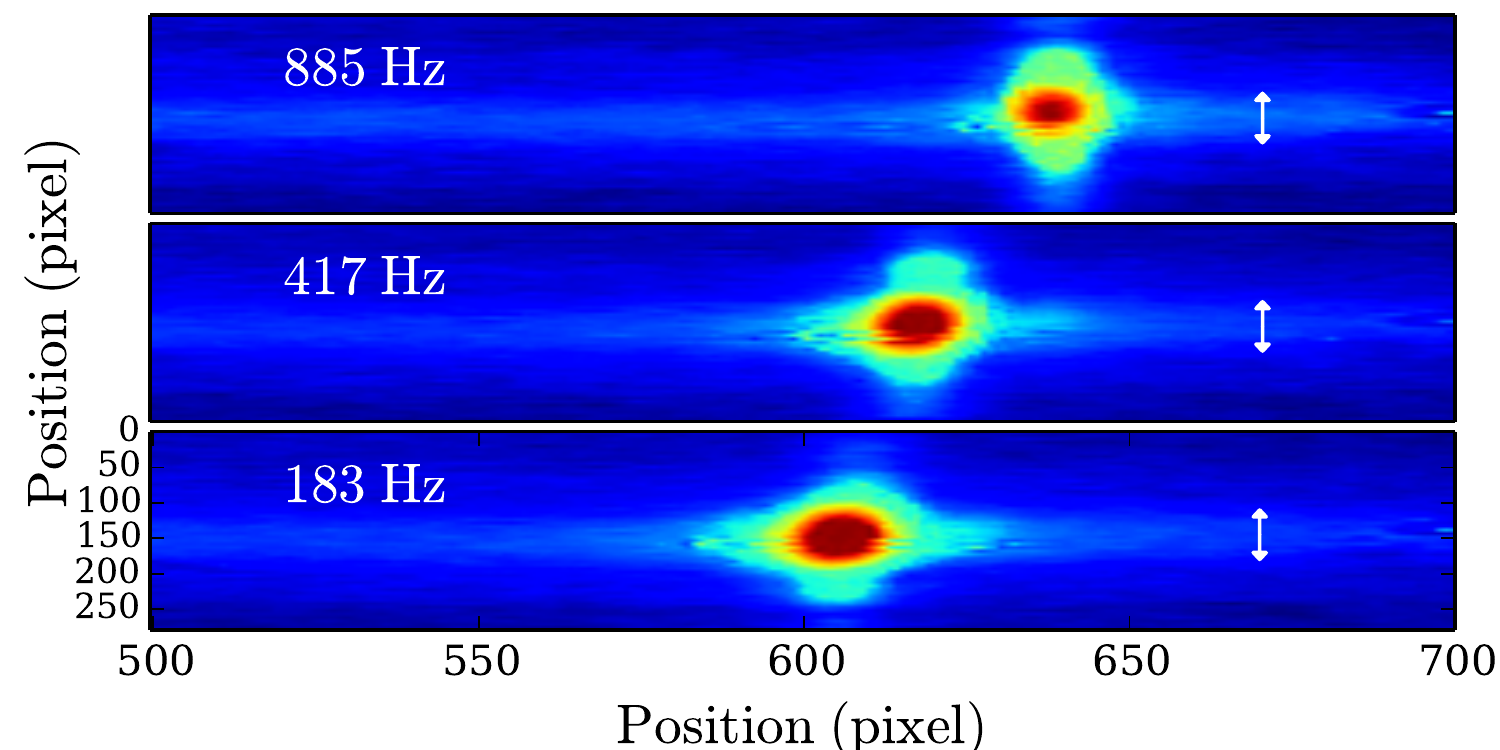}

\caption{Typical camera images in Setup 1 at rotation frequencies 183, 417,
and 885\,Hz, respectively. The profiles below were obtained by integrating
over the region indicated on the right hand side by the small white
arrows. }

\label{fig:camera_image}
\end{figure}

Having determined the calibration for both the camera, and the rotation
speeds, we now turn to the measurement of the displacements. First
we discuss the results obtained from Setup 1. Typical images of the
reflected beam at three different rotation speeds (183, 417, and 885\,Hz)
are shown in Fig.\,\ref{fig:camera_image} (only a small part of
the otherwise 720-by-1280 chip is displayed). The movement of the
beam is clearly visible. Note that, while we begin with a circularly
symmetric Gaussian beam (this is what leaves the single-mode fiber),
the camera image is elongated along the vertical direction, which
is perpendicular to the direction of the displacement. The reason
for this is that the mirrors are only 2\,mm thick, but 15\,mm wide,
while the beam at the mirror's position is still about 10\,mm in
diameter. This means that diffraction will stretch the beam in the
direction of the smallest dimension of the mirror. 

The images in Fig.\,\ref{fig:camera_image} are turned into nearly
Gaussian profiles by vertically integrating over a range of $\pm25$
pixels around the maximum, as indicated by the small white arrows
in the figure. Such profiles for three different rotation speeds (183,
417, and 885\,Hz) are shown in Fig.\,\ref{fig:profiles}. In order
to accurately determine the centre positions of these profiles, we
fit a Gaussian with an offset to the data points in a range of $\pm15$
pixels around the pixel with the highest intensity, as shown by the
shaded gray domains in the figure. The centre of these fits is then
accepted as the true position of the reflected beam. The error in
the fit is less than $0.15$ pixels for all measurements. 

\begin{figure}[h]
\includegraphics[width=1\columnwidth]{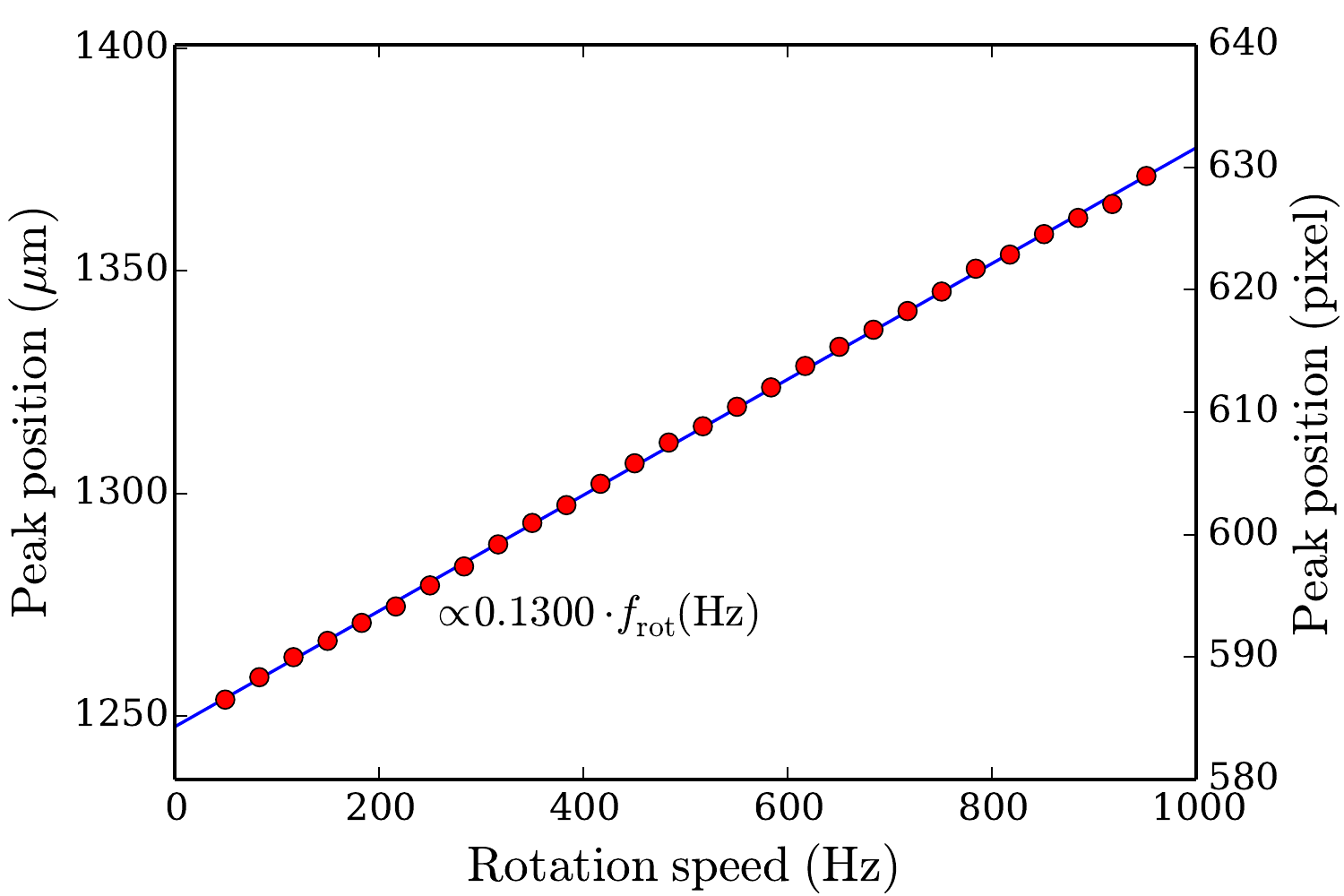}

\caption{Position of the reflected beam as a function of the rotation speed
in Setup 1. On the right vertical axis, the same data are shown in
units of the CCD pixels. The peak position can be obtained from $P_{\mathrm{peak}}(\mathrm{\mu m})=(0.130\pm0.00047)\cdot f_{\mathrm{rot}}\mathrm{(Hz)}+1247.6(\mathrm{\mu m)}$.
Error on the data points is not visible.}

\label{fig:shift_vs_frequency}
\end{figure}

Figure\,\ref{fig:shift_vs_frequency} contains measurement data on
the beam displacement as a function of the rotation speed. On the
right vertical axis, the positions are given in terms of the CCD pixels,
as taken from images similar to Fig.\,\ref{fig:camera_image}. The
left axis displays the positions in physical units, after the CCD
pixels were converted using the fit from Fig.\,\ref{fig:ccd_calibration}.
The linear fit to these data yields a slope of $(0.130\pm0.00047)\,\mathrm{\mu m/Hz}$.
Given that, with the nomenclature of Eq.\,(\ref{eq:c_diffs}), $d_{1}=425\pm1\,\mathrm{mm}$,
$d_{2}=1630\pm1\,\mathrm{mm}$, and $d_{3}=4830\pm1\,\mathrm{mm}$,
and taking all above-mentioned error sources into account, we calculate
a speed of light of $c=(2.97\pm0.03)\cdot10^{8}\,\mathrm{m/s}$. This
is within 1\% of the defined value of $2.99792458\cdot10^{8}\,\mathrm{m/s}$,
and overall, the statistical errors are within 1\%.

\begin{figure}[h]
\includegraphics[width=1\columnwidth]{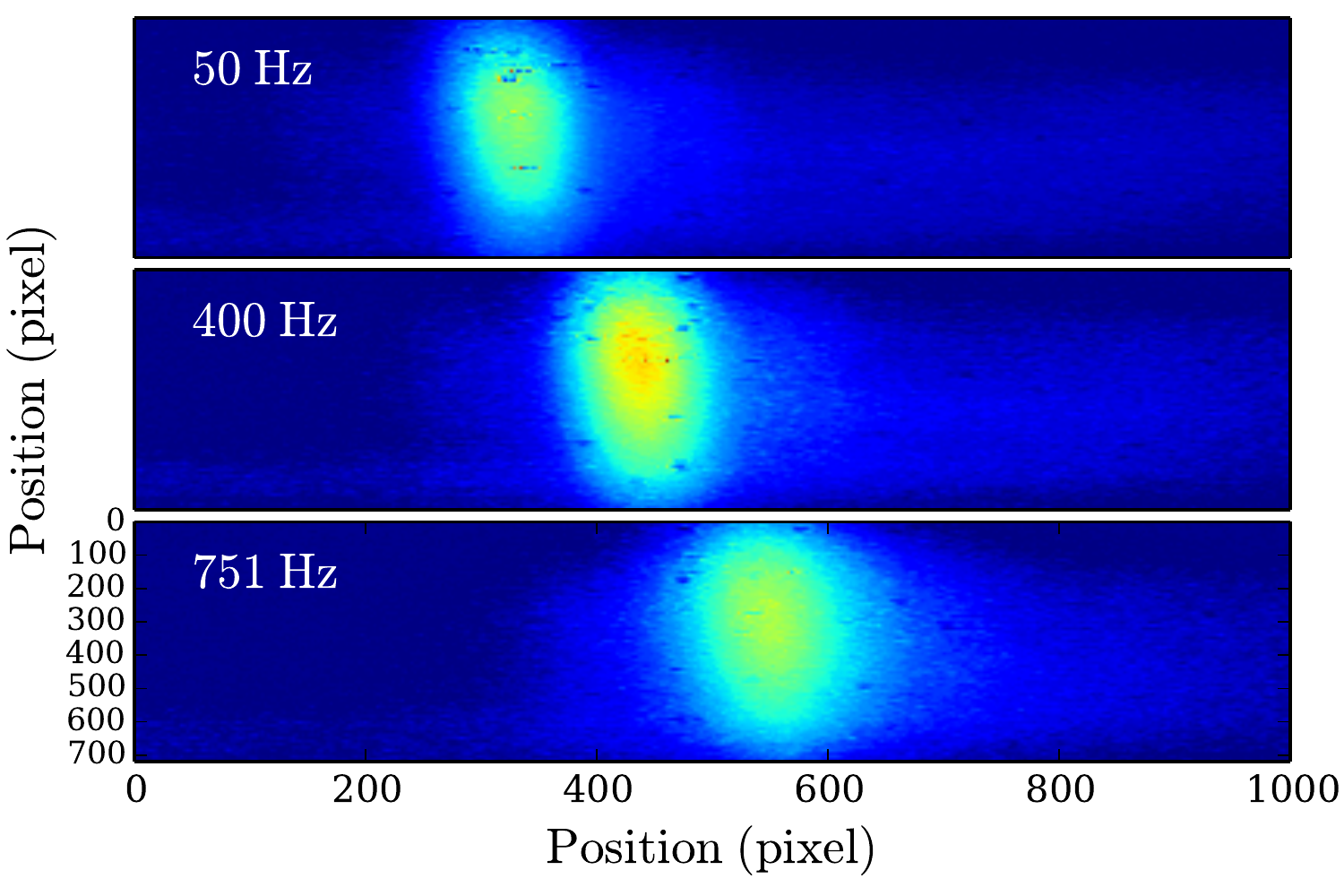}

\caption{Camera images in Setup 2 at frequencies 50, 400, and 751\,Hz.}

\label{fig:camera_image_setup2}
\end{figure}

We now discuss measurements in Setup 2. Typical camera images at frequencies
50, 400, and 751\,Hz, respectively are shown in Fig.\,\ref{fig:camera_image_setup2}.
As opposed to the other setup, the laser spot is stretched vertically
over the whole length of the camera (720 pixels). Also note that as
the frequency increases, so does the width of the images. We speculate
that this might be related to turbulence generated by the fast rotating
mirrors: while the average speed of the motor is determined by the
clock frequency, vortices detaching from the vertices of the octagonal
mirror can lead to fluctuations in the instantaneous speed.

\begin{figure}[h]
\includegraphics[width=1\columnwidth]{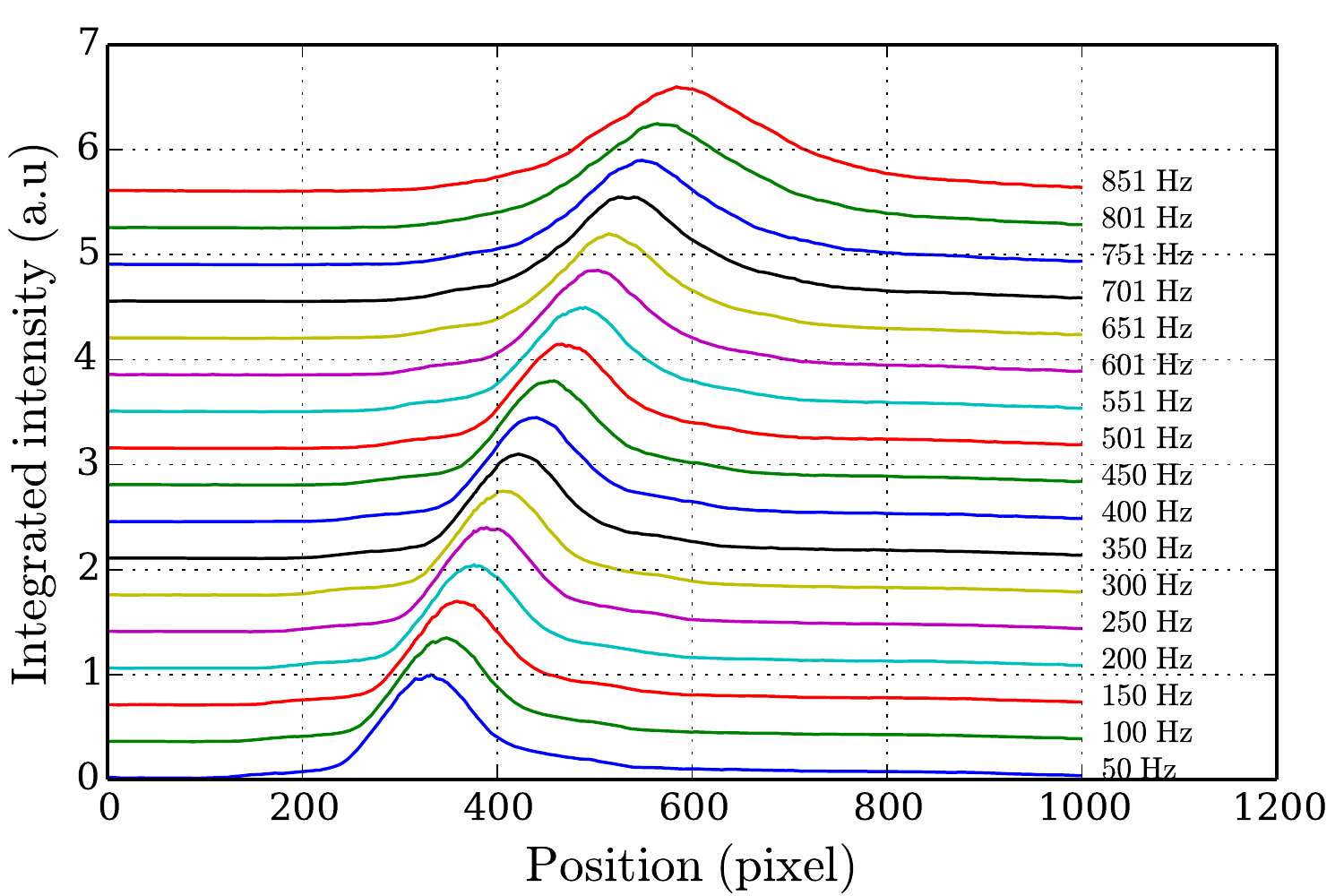}

\caption{Vertically integrated camera profiles in Setup 2 as a function of
the frequency.}

\label{fig:sep10_set3_profiles}
\end{figure}

This change in the width can also be seen in Fig.\,\ref{fig:sep10_set3_profiles},
where we plot the vertically integrated camera images for 17 rotation
frequencies as indicated. However, despite the broadening of the profiles,
the displacement is clearly visible as the frequency changes.

\begin{figure}[h]
\includegraphics[width=0.98\columnwidth]{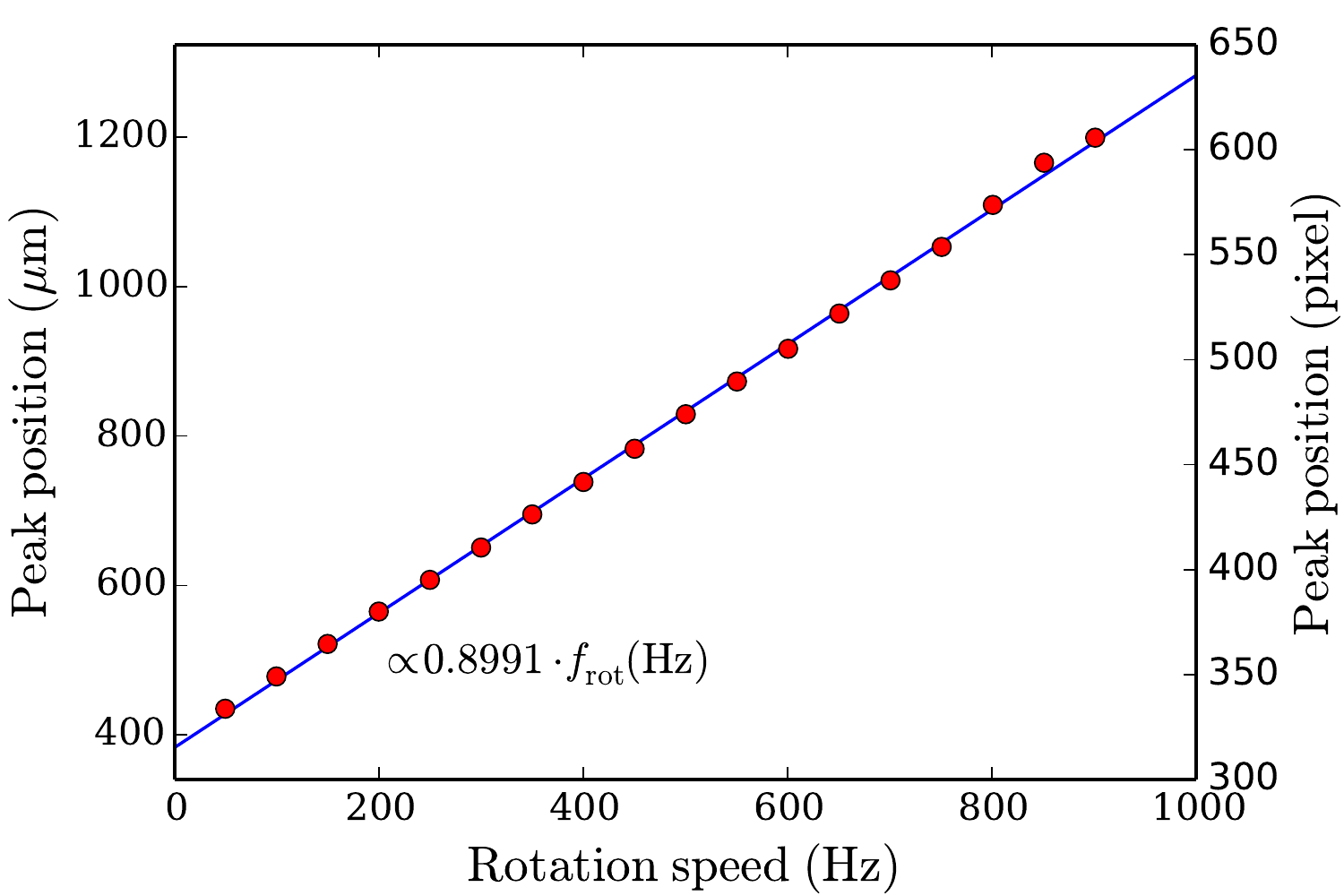}

\caption{Position of the reflected beam as a function of the rotation speed
in Setup 2. On the right vertical axis, the same data are shown in
units of the CCD pixels. The linear fit to the peak position is $P_{\mathrm{peak}}(\mathrm{\mu m})=(0.899\pm0.0059)\cdot f_{\mathrm{rot}}\mathrm{(Hz)}+383.3(\mathrm{\mu m)}$.
Error on the data points is not visible.}

\label{fig:shift_vs_freq_sep10}
\end{figure}

In Fig.\,\ref{fig:shift_vs_freq_sep10} we plot the beam displacement
as a function of the rotation speed, similar to Fig.\ref{fig:shift_vs_frequency}.
The linear fit to these data yields a slope of $(0.899\pm0.0059)\,\mathrm{\mu m/Hz}$.
Given that $d_{1}=4060\pm1\,\mathrm{mm}$, $d_{2}=730\pm1\,\mathrm{mm}$,
and $d_{3}=3260\pm1\,\mathrm{mm}$, and considering all error sources,
we calculate a speed of light of $c=(3.02\pm0.03)\cdot10^{8}\,\mathrm{m/s}$.

Our experimental conditions and results are summarized in Table\,\ref{table:experimental_conditions}.

\begin{table}[h]
\begin{tabular}{|c|c|c|c|c|}
\hline 
 & $d_{1}$ & $d_{2}$ & $d_{3}$ & c (m/s)\tabularnewline
\hline 
\hline 
\multirow{2}{*}{Setup 1} & $\overline{FP-L_{2}}$ & $\overline{L_{2}-RM}$ & $\overline{RM-SM}$ & \multirow{2}{*}{$2.97\cdot10^{8}$}\tabularnewline
\cline{2-4} 
 & 425\,mm & 1630\,mm & 4830\,mm & \tabularnewline
\hline 
\multirow{2}{*}{Setup 2} & $\overline{FP-SM}$ & $\overline{SM-RM}$ & $\overline{RM-FM}$ & \multirow{2}{*}{$3.02\cdot10^{8}$}\tabularnewline
\cline{2-4} 
 & 4060\,mm & 730\,mm & 3260\,mm & \tabularnewline
\hline 
\end{tabular}

\caption{Summary of experimental conditions, and results. Overlines denote
distances between designated elements.}

\label{table:experimental_conditions}
\end{table}

\section{Systematic errors}

We have already indicated the magnitude of statistical errors: the
calibration of the CCD is about 0.2\%, the rotation frequency's is
about 0.3\%, the length measurement's is less than 0.1\%, while the
Gaussian fits to the profiles contain an error of about 0.2\%. However,
in addition to these, there are a number of systematical errors that
one has to consider. 

One we have already pointed out, namely, if the camera is not perpendicular
to the laser beam, all displacements will be measured \emph{shorter}
and this will lead to a seemingly \emph{larger }speed of light. One
way of removing this error source is to slightly rotate the camera
without moving it, and repeat the measurements multiple times. The
smallest value of $c$ should correspond to the perpendicular configuration.
However, since this correction is proportional to the cosine of the
angle of deviation from the normal, errors are of second order.

Second, if the camera's plane is not parallel to the axis of the translation
stage, the pixel size will be inferred incorrectly, and this, again,
will lead to a seemingly higher light speed. As mentioned above, a
trivial test for this is the beam profile measured at various positions
of the translation stage: all other conditions being identical, a
simple translation should result in identical profiles. If this is
not the case, then the camera has to be rotated slightly with respect
to the translation stage till all measured profiles are identical.
As with the systematic error discussed above, corrections are quadratic
in the angle. 

Third, the measurement of independent quantities, in this case, the
frequency (time) and distance might contain errors that result from
the particular method used to measure them. Given the accuracy of
frequency measurements, it is reasonable to expect that only the value
of distance would be affected, and one can safely neglect systematic
errors in frequency.

\begin{figure}[h]
\includegraphics[width=0.98\columnwidth]{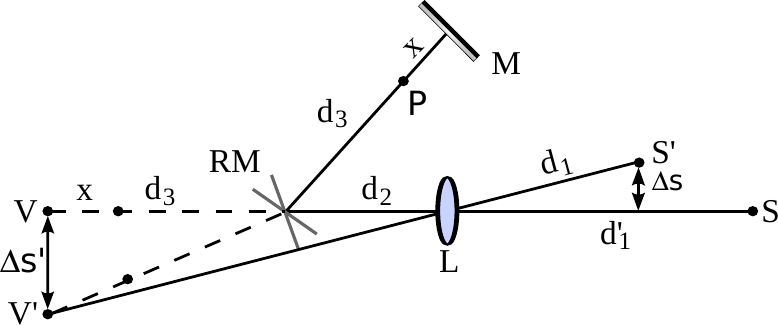}

\caption{The concept of the measurements, with focusing errors. Notation as
in Fig.\,\ref{fig:concept}. $P$ is the image of the source $S$.}

\label{fig:concept_imperfect}
\end{figure}

Fourth, imperfections in the focusing lead to small errors. In order
to estimate the order of magnitude of these, let us assume that the
image of $S$ is formed not at the end mirror, but at $P$, which
is at a distance of $x$ from $M$, as shown in Fig.\,\ref{fig:concept_imperfect}.
The virtual image of $P$ will also be shifted by the same amount,
and following the derivation in Section III., we arrive at 
\begin{equation}
c=\frac{4d_{1}d_{3}(d_{3}-x)}{d_{2}+d_{3}-x}\left(\frac{d\Delta s}{d\omega}\right)^{-1}\approx c_{0}\left[1-\frac{xd_{2}}{d_{3}(d_{2}+d_{3})}\right]\ ,\label{eq:c_diffs_imperfect}
\end{equation}
if $x\ll d_{3}$. Note that $d_{1}$ does not necessarily indicate
the distance at which the image is formed: it simply designates the
position of the measurement (webcam). If $d_{1}$ is chosen such that
the imaging condition is not satisfied, it does not mean that the
derivation is incorrect, it only means that the image will not be
sharp at that point, but Eq.(\ref{eq:c_diffs_imperfect}) is still
valid. 

The magnitude of the correction will depend on two parameters of the
setup, $d_{2},d_{3}$, and the inaccuracy in the focusing, $x$. Note
that for $d_{2}=0$, i.e., when the rotating mirror is next to the
imaging element, the first-oder correction is zero. In the first setup
$d_{2}/d_{3}\approx1/3$, while in the second case, $d_{2}/d_{3}\approx1/4$.
Therefore, an upper bound for the correction in Eq.(\ref{eq:c_diffs_imperfect})
is $x/(4d_{3})$. Given that $d_{3}\geq3\,\mathrm{m}$, we incur an
error of 1\%, if $x\approx0.12\,\mathrm{m}$. It is reasonable to
assume that the focus can be determined with $10\,\mathrm{cm}$ accuracy,
even if the imaging elements have such long focal length. Therefore,
we can conclude that the error related to imperfect focusing is less
than 1\%.

Finally, the lens, the only glass element in the first setup, has
finite width with a refractive index larger than one, and this adds
to the total length between the focal point and the end mirror. This
extra optical length can be measured and added to the path, provided
the refractive index of the glass is known. Of course, the second
setup does not suffer from this kind of error.

\section{Conclusion}

In conclusion, we presented a simple version of the Foucault method
for the measurement of the speed of light. We demonstrated that with
readily available and inexpensive optics, and a bit of data processing,
acceptable accuracy (results within 1\% of the defined value) can
be achieved. We also discussed a range of systematic errors, and pointed
out several possible improvements. The experiment teaches students
the historically important Foucault method, and modern data evaluation
concepts at the same time. 

\bibliographystyle{apsrev}
\bibliography{speedoflight}

\appendix

\section{MATLAB code}

Here we list matlab snippets that can be used for the evaluation of
images. The usual workflow is to create a profile similar to that
in Fig.\,\ref{fig:profiles} with the function \texttt{create\_profile},\emph{
}and pass the output to the function \texttt{fit\_profile},\emph{
}which will print the parameters of the best Gaussian fit to the console.
\texttt{gauss} simply defines the fit function, and it can easily
be replaced by other, more appropriate forms, if necessary.

\begin{widetext}

\inputencoding{latin9}\begin{lstlisting}
function profile = create_profile(fn, range1, range2) 
	% Returns and displays a profile (a vertically integrated image segment)
	% 
	% Input:
	%
	% fn: the file (camera image) to read
	% range1, range2: the limits of the vertical intagration

	im = imread(fn); 
	im1 = sum(im, 3); % Turns the RGB image into grayscale
	figure(1); imagesc(im1)
	profile = sum(im1(range1:range2,:)); 
	figure(2); plot(profile, 'ro'); 
\end{lstlisting}
\inputencoding{utf8}

\inputencoding{latin9}\begin{lstlisting}
function y = gauss(par, xdata) 
	% Returns a Gaussian function evaluated at points given in xdata
	% 
	% y = A*exp(-(x-x0)^2/sigma^2)+offset 
	% 
	% Input:
	% par: an array of parameters in the form [A, x0, sigma, offset]
	% A: amplitude of the Gaussian
	% x0: centre of the Gaussian
	% sigma: standard deviation of the Gaussian
	% offset: offset of the Gaussian
	% xdata: points where the function is to be evaluated

	y = par(1) * exp(-(xdata-par(2)).^2/par(3)/par(3)) + par(4);
\end{lstlisting}
\inputencoding{utf8}

\inputencoding{latin9}\begin{lstlisting}
function fit_profile(profile, xlim1, xlim2, par) 
	% Fits a profile with a Gaussian
	%
	% Input:
	% profile: a 1D array of evenly spaced measurement points
	% xlim1, xlim2: the fit will be done on domain [xlim1:xlim2]
	% par: initial guess of the fit (see gauss(par, xdata))

	f = lsqcurvefit(@gauss, par, xlim1:xlim2, profile(xlim1:xlim2)) 
	figure(3); hold on; plot(xlim1:xlim2, profile(xlim1:xlim2), 'ro')
	plot(xlim1:xlim2, gauss(f, xlim1:xlim2), 'b-');
	hold off;
\end{lstlisting}
\inputencoding{utf8}

\end{widetext}
\end{document}